\documentclass[conference]{IEEEtran}
\IEEEoverridecommandlockouts
\usepackage{cite}
\usepackage{amsmath,amssymb,amsfonts}
\usepackage{algorithmic}
\usepackage{graphicx}
\usepackage{textcomp}
\usepackage{xcolor}
\usepackage{multirow}
\usepackage{graphicx}

\usepackage[utf8]{inputenc}
\usepackage{booktabs}
\usepackage{subfig}
\usepackage{xspace}
\usepackage{siunitx}

\makeatother
\def\BibTeX{{\rm B\kern-.05em{\sc i\kern-.025em b}\kern-.08em
    T\kern-.1667em\lower.7ex\hbox{E}\kern-.125emX}}

\usepackage{amssymb}
\usepackage{pifont}
\newcommand{\pb}[1]{\vspace{0.75ex}\noindent{\bf \em #1}\hspace*{.3em}}

\DeclareMathSymbol{*}{\mathbin}{symbols}{"03} 
\DeclareMathSymbol{\ast}{\mathbin}{symbols}{"03}

\begin{document}









\title{How Impersonators Exploit Instagram \\to Generate Fake Engagement?}



\vspace{-0.4cm}

\author{\IEEEauthorblockN{Koosha Zarei\IEEEauthorrefmark{1}, Reza Farahbakhsh\IEEEauthorrefmark{1}, No\"{e}l Crespi\IEEEauthorrefmark{1}}
\IEEEauthorblockA{\IEEEauthorrefmark{1}\textit{CNRS Lab UMR5157,  T\'el\'ecom SudParis, Institut Polytechnique de Paris,} \textit{Evry, France}.\\
\{koosha.zarei, reza.farahbakhsh, noel.crespi\}@telecom-sudparis.eu}
}
\vspace{-0.4cm}

\maketitle

\begin{abstract}
Impersonators on Online Social Networks such as Instagram are playing an important role in the propagation of the content. These entities are the type of nefarious fake accounts that intend to disguise a legitimate account by making similar profiles. In addition to having impersonated profiles, we observed a considerable engagement from these entities to the published posts of verified accounts. Toward that end, we concentrate on the engagement of impersonators in terms of active and passive engagements which is studied in three major communities including ``Politician'', ``News agency'', and ``Sports star'' on Instagram. Inside each community, four verified accounts have been selected. Based on implemented approach in our previous studies \cite{Zare1910:Typification}, we have collected 4.8K comments, and 2.6K likes across 566 posts created from 3.8K impersonators during 7 months. Our study shed light into this interesting phenomena and provides a surprising observation that can help us to understand better how impersonators engaging themselves inside Instagram in terms of writing Comments and leaving Likes.

\end{abstract}


\section{Introduction}

Impersonators are most commonly found on all major social media platforms including Facebook, Twitter, Instagram, YouTube and LinkedIn. Among these platforms, Instagram is widely used by celebrities and influencers with different level of popularity and visibility for their everyday activities and news propagation. On other hand, this is a golden opportunity for impersonators and recent studies show their considerable presence in Instagram. 
In this era, fake news and weaponized information is still a very hot topic and as it is presented in previous studies \cite{kooshaDeep} \cite{Zare1910:Typification}, one of the most common ways of spreading fake news, disinformation, or false activities is using fake profiles, where malicious users create social media accounts impersonating 
a legitimate account and present themselves in profiles who are very similar to real persons in term of profile metrics. This activity named as ``Impersonating" and impersonators are those accounts that are pretending to be someone well-known or representative of a known brands, company etc.

Furthermore, Fake Instagram profiles have a pretty clear plan—they make accounts appear more popular than they are and with bot services, they create fake engagement, too. Fake engagements strike social media and especially Instagram, which makes it considerably harder to understand which posts are genuinely getting the best reaction from legitimate accounts/followers.

From malicious activities in social media, a larger set of threats has been identified including brand abuse, fraud and follower farming. Therefore several lawsuits has been taken in place in United State (along with other countries), where criminal impersonation is a crime that is governed by states laws, which vary by state. It involved assuming a false identity with the intent to defraud another or pretending to be a representative of another person or organisation \cite{impersonaion_law}.

In this paper, we aim to understand the impact of impersonators on fake content production and propagation by analysing the tactics meant to lure the user attractions to the produced or fabricated content. Toward that end, we picked three distinct communities on Instagram including ``Politician'', ``News agency'', and ``Sports star''. Inside each category, we selected four top verified genuine accounts and we collected a great number of posts beside comments and likes (in a 7-month period). For each one, we detected and extracted the impersonator profiles based on our methodology presented in \cite{Zare1910:Typification}, and we ended up with 3.8K dataset. Next, we clustered them into three groups \textit{C0-Fan-Pages}, \textit{C1-Ordinary-Users}, and \textit{C2-BotLike} based on profile and activity characteristics. 

In this study, We first, investigate the portion of the comment, like, and post that are distributed by impersonators across communities. Next, to understand what is being shared, we analyse the comments. In this regard, we use natural language processing (NLP) techniques to understand the context of the written text and analyse the semantic and sentiment aspects. The contribution of this study can be summarised as follow:

\begin{itemize}
\item We assemble a precious dataset of the content and activities of impersonators in three leading communities.

\item provides a comprehensive analysis of the behaviour of impersonators in the shape of active and passive engagement.

\item provides the first analysis of how impersonators create fake engagements across leading communities on Instagram.

\item presents an investigation of the content that is produced by impersonators across communities which potentially lead us to type of fake contents.

\end{itemize}

The remaining of this study is as follows. Section \ref{related_work} gives the related studies. The process of data crawling, the description of communities, validation and the dataset are described in section \ref{data_collection}. The concept of detection of impersonators is specified in \ref{identification}. Next, we investigate the behaviour of impersonators and the communities they target in section \ref{behaviour_of_impersonator}. Next, we analyse the content that is distributed by impersonators in section \ref{fake_content}. Finally, section \ref{future} shows future directions and concludes the study.

\section{Related work}
\label{related_work}

\pb{Fake account:} Recent research has worked on related research problems and dedicated a fair amount of work to study a different aspect of OSNs. In this era, looking to behavioural aspect of users and understand the different pattern of activities is still a hot topic of research. Several studies tried to shed light on this direction by profiling users based on their activities and reactions. This work \cite{8622011} presents a novel technique to discriminate real accounts on social networks from fake ones. The writers from this \cite{RAMALINGAM2018165} study provide a review of existing and state-of-the-art Sybil detection methods with an introductory approach and present some of the emerging open issues for Sybil detection in Online Social Networks.

\pb{Bot:} On the other hand, the huge existence of Bots can alter the perception of social media influence, artificially enlarging the audience of some people, or they can impact the reputation of a company. The problem of rising social bots are discussed in \cite{Ferrara:2016:RSB:2963119.2818717}. There are various strategies to tackle the problem of bot detection. \cite{profchar} suggested a profile-based approach and \cite{7865975} proposed a novel framework on detecting spam content. Also, \cite{Xiao:2015:DCF:2808769.2808779} presented a machine learning pipeline for detecting fake accounts and authors in \cite{stweeler, tweb19} present a method to classify bots and understand their behaviour in scale.

\pb{Fake Engagement:}. From this viewpoint, Authors in \cite{Li_2016_WCC}, focus on the social site of YouTube and the problem of identifying bad actors posting inorganic contents and inflating the count of social engagement metrics. They propose an effective method and show how fake engagement activities on YouTube can be tracked over time. Likewise, another study, \cite{Sen:2018:WWL:3201064.3201105}, enumerate the potential factors which contribute towards a genuine like on Instagram. Based on analysis of liking behaviour, they build an automated mechanism to detect fake likes on Instagram which achieves a high precision of 83.5

\pb{User Behaviour:} On another line of research, the authors in \cite{Buccafurri:2015:CTF:2822539.2822620} \cite{Lim:2015:MVI:2808797.2808820} look at the profile and behavioural patterns of a user and discussed existing challenges on different OSNs. By integrating semantic similarity and existing relationships between users, it is possible to match profiles across various OSNs \cite{Choumane:2017:PMS:3093241.3093258}  \cite{Krombholz2012}. 
Also, \cite{Goga:2015:RPM:2783258.2788601} conducted a detailed investigation of user profiles and proposed a matching scheme. On Instagram, for the sake of mitigating impersonation attack, \cite{Sen:2018:WWL:3201064.3201105} explored fake behaviours and built an automated mechanism to detect fake activities. 

As far as our best knowledge, the problem of spotting and analysing the fake engagement is not studied in the literature and this is the first study that analyzed this phenomenon through the lens of impersonators on Instagram.


\section{Data Collection}\label{data_collection}
Considering the Instagram API policies, we implemented an exclusive crawler in Python to receive data and store in a MongoDB server in the form of JSON files. We use the official Instagram API \cite{InstagramAPI:online} which is based on the Facebook Platform to gather all posts, comments, and likes. This will return posts concerning Instagram rules. Note that we only gather public data excluding any potential sensitive data. The whole data collection process is designed exclusively for research purposes and the data is stored in an anonymized format.

\subsection{Communities and Case Studies}\label{data_communities}
To investigate and understand the behaviour of impersonators, it is essential to have a dataset that consists of data from a variety of categories. Toward that end, we examined impersonators in three influential communities including politician, news agencies and sports stars. As a result, we are dealing with a wide range of profile characteristics and user behaviours. In such a scenario, we have targeted the top famous figures inside each community. All genuine accounts are official pages, have \textit{Verified} Badge and are confirmed by Instagram \cite{InstagramBadge}.
Next we explain briefly each category and the target users inside each category in this study.

\begin{itemize}

\item \textbf{Politician} community is of high interest. Having a large number of followers, fan pages, oppositions and supporters are the main reasons for selecting this community. Additionally, Political Bot is a new phenomenon in this area. Donald J. Trump (\textit{@realdonaldtrump}) the president of the United States, Barack Obama (\textit{@barackobama}) the previous president of the United States, Emmanuel Macron (\textit{@emmanuelmacron}) the president of France, and Theresa May (\textit{@theresamay}) the Prime Minister of the United Kingdom (all at the time of writing this paper) are included as target users in our dataset.

\item \textbf{News Agency} is another vital community in which top English language news broadcasters including BBC (\textit{@bbc}),  CNN (\textit{@cnn}),  FoxNews (\textit{@foxnews}), and Reuters (\textit{@reuters}) are considered. Use of Social Media is changing the relationship between the news agencies and the audience. This community has a large number of followers from various groups which make it very interesting category for the popuse of this study.

\item \textbf{Sports Star} community represents top sports players in football and tennis. Nowadays, thanks to social media, we see sporting star's habits, milestones and personal lives every day on our phones. Fake news, Fake profiles, and Disinformation are considered as serious difficulties inside this community. Leo Messi (\textit{@leomessi}), Cristiano Ronaldo (\textit{@cristiano}), Rafael Nadal (\textit{@rafaelnadal}), and Roger Federer (\textit{@rogerfederer}) are selected.

\end{itemize}


\subsection{Dataset} \label{dataset_subsection}

In this study, we use the dataset which is obtained from our previous studies \cite{kooshaDeep} \cite{Zare1910:Typification} and the primary target is to analyse the content that is generated by the impersonators and investigate the fake engagement. First, we target the previously mentioned well-known figures on Instagram (see \ref{data_communities}) and collect their activity from October 2018 until April 2019. The activity includes posts, comments, likes, and user information. Based on our methodology, from the the pool of users who reacted in the shape of comment and like, we extract and identified 3.8K unique impersonators. Next, based on different metrics, we clustered impersonators into three main clusters (for more details please see \ref{identification}). 
In total, our dataset includes 3.8K impersonators who generate 4.8K comments and 2.6K likes across 566 unique posts during the period of 7 months.

\subsection{Validation}
A natural risk is that a subset of the comment and likes that are given to posts may be generated by users who are not impersonators. So, we further perform manual annotation to validate the general correctness of our data. To validate our dataset, we manually looked at the profiles of the impersonators to verify if they were really impersonator. To validate profiles, accounts of three clusters are completely checked manually. We filter any incorrectly identified impersonators.

\pb{Ethics}: In line with Instagram policies, user privacy and ethical consideration defined by the community, we only gather publicly available data that are obtainable from Instagram.

\section{Who are Impersonators?}\label{identification}

\pb{Phase 1: Impersonator Detection.} An impersonator is someone who pretends or copies the behaviour or actions of another. Of course, there are many reasons for impersonating someone. In the first study \cite{kooshaDeep} we answered questions like who are the impersonators? What is the rate of engagement in the shape of like and comment? How many impersonators exist? and What is the activity of this group? We studied politician community with 3 use cases (D. Trump, B. Obama, E. Macron) on Instagram and track their activity (with user reactions) for three months. We presented a methodology to detect impersonators based on the profile similarity and we discovered more than 200 fake accounts with different levels of similarity. Interestingly, While Trump held the most impersonators, but Macron contained the least (108 vs. 21).

\begin{table}[ht]
\vspace{-0.1cm}
\caption{Summary of Impersonators across Clusters}
\begin{center}
\vspace{-0.3cm}
\resizebox{\columnwidth}{!}{
\begin{tabular}{| c | c | c | c | c | c |}
\hline
\textbf{Clusters} & \textbf{Type} & \textbf{\begin{tabular}[c]{@{}c@{}}\#of unique\\ account\end{tabular}} & \textbf{\#of comment} & \textbf{\#of like} & \textbf{\#of post$^{\mathrm{\textbf{*}}}$} \\ \hline
C0\_Fan\_Page & Fan Page & 54\% & 52\% & 50\% & 36\% \\ \hline
C1\_Ordinary\_User & Normal User & 34\% & 37\% & 29\% & 24\% \\ \hline
C2\_Botlike & Bot & 12\% & 11\% & 19.4\% & 40\% \\ \hline
\multicolumn{2}{| c |}{\textbf{Total Number}} & \textbf{3.8K} & \textbf{4.8K} & \textbf{2.6K} & \textbf{566}  \\ \hline
\multicolumn{4}{ c }
{$^{\mathrm{\textbf{*}}}$the number of unique posts which impersonators reacted to.}
\end{tabular}
}
\end{center}
\label{table_dataset_cluster}
\vspace{-0.2cm}
\end{table}

\pb{Phase 2: Clustering.} We, next in the second study \cite{Zare1910:Typification} investigated that impersonators are more interested in which community? Among them, how many distinct hidden groups exist? what are their characteristics? and how impersonators are involved in terms of reactions? To answer these questions as we extended the dataset to 3 communities including `politician', `news agency', and `sports star' communities with 12 famous verified use cases (see \ref{data_communities}), also we enhanced the detection methodology. So we ended up with 3.8k impersonators with various characteristics. We, next applied three major clustering methods including K-means, Gaussian Mixture Model, and Spectral Clustering algorithms. We divided impersonators into 3 clusters based on 10 features such as \textit{username similarity}, \textit{name similarity}, \textit{bio similarity}, \textit{photo similarity}, \textit{most common metrics (mcm)}, \textit{number of followers}, \textit{number of followees}, \textit{number of media count}, \textit{private status}, and \textit{verified status}. Next, Based on their characteristics and behaviours we are calling them \textit{C0-Fan-Pages}, \textit{C1-Ordinary-Users}, and \textit{C2-BotLike}. Table~\ref{table_dataset_cluster} summarises the dataset across clusters. Rows present clusters and columns present data types.

\begin{figure}[ht]
\vspace{-0.3cm}
  \subfloat[Active Engagement / community]{\includegraphics[width=0.49\linewidth, , height=0.20\textwidth]{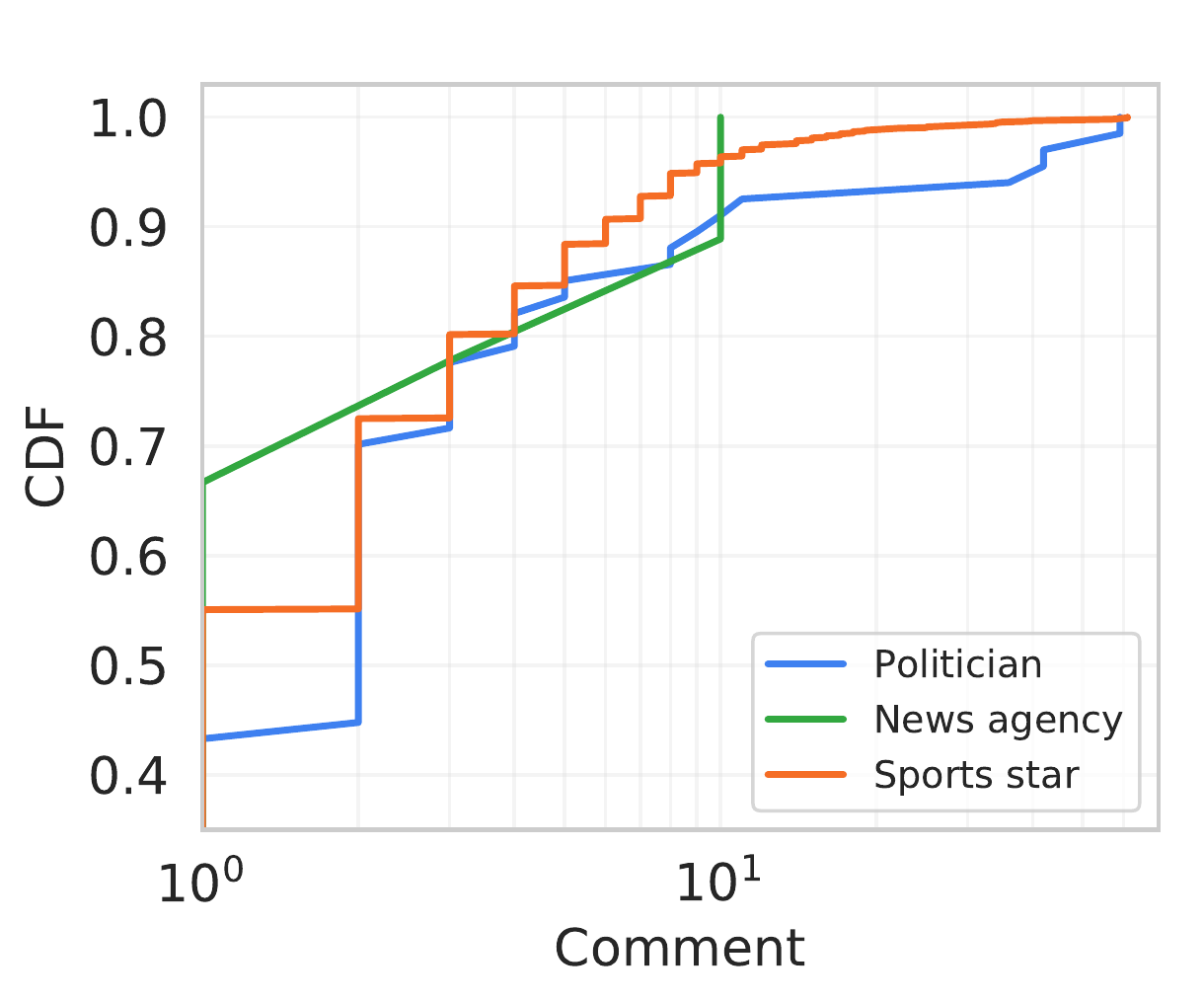}}\hfill
  \subfloat[Active Engagement / cluster]{\includegraphics[width=0.49\linewidth, height=0.20\textwidth]{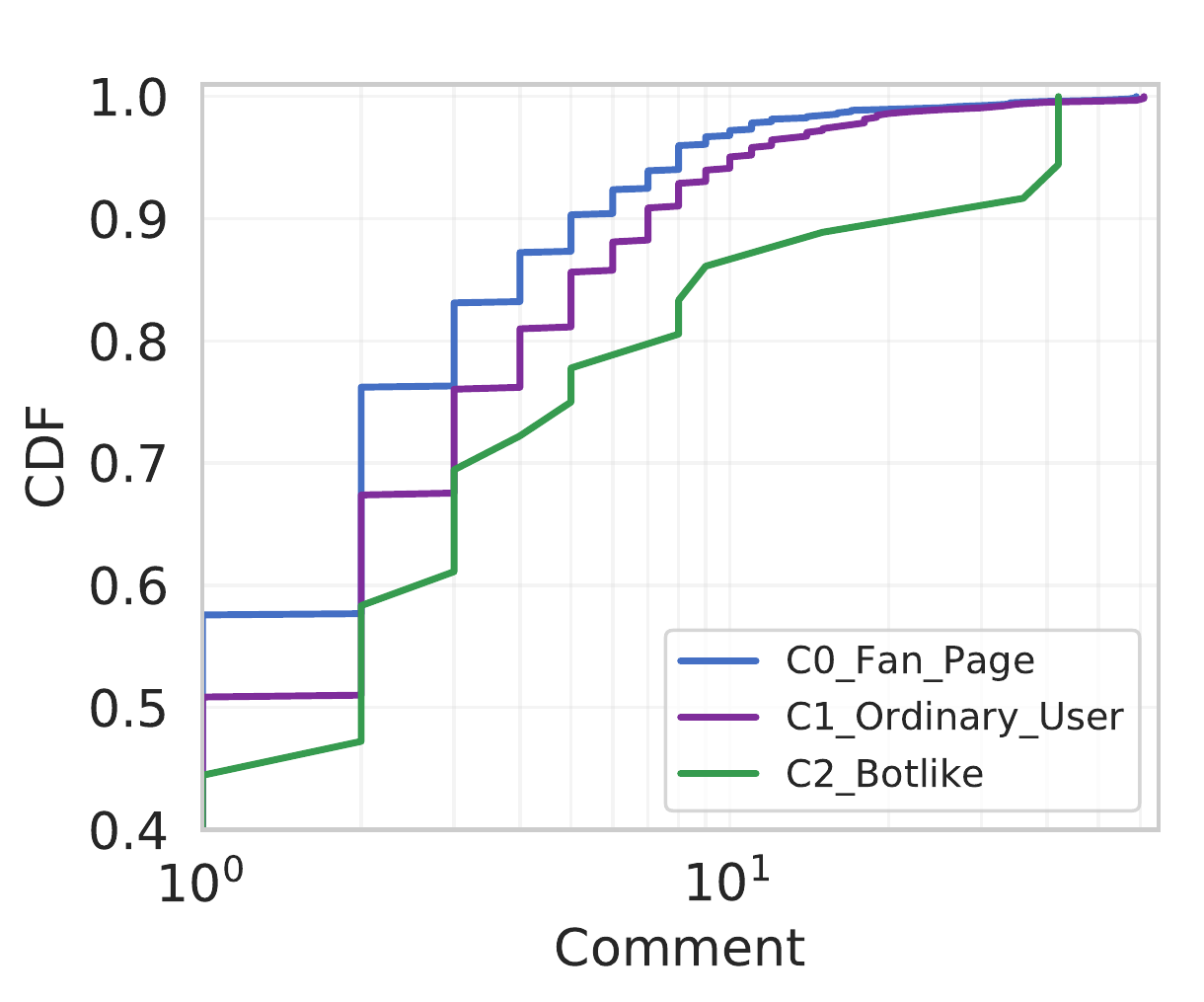}}%
 
  \vspace{-0.2cm}
 
  \subfloat[Passive Engagement / community]{\includegraphics[width=0.49\linewidth, , height=0.20\textwidth]{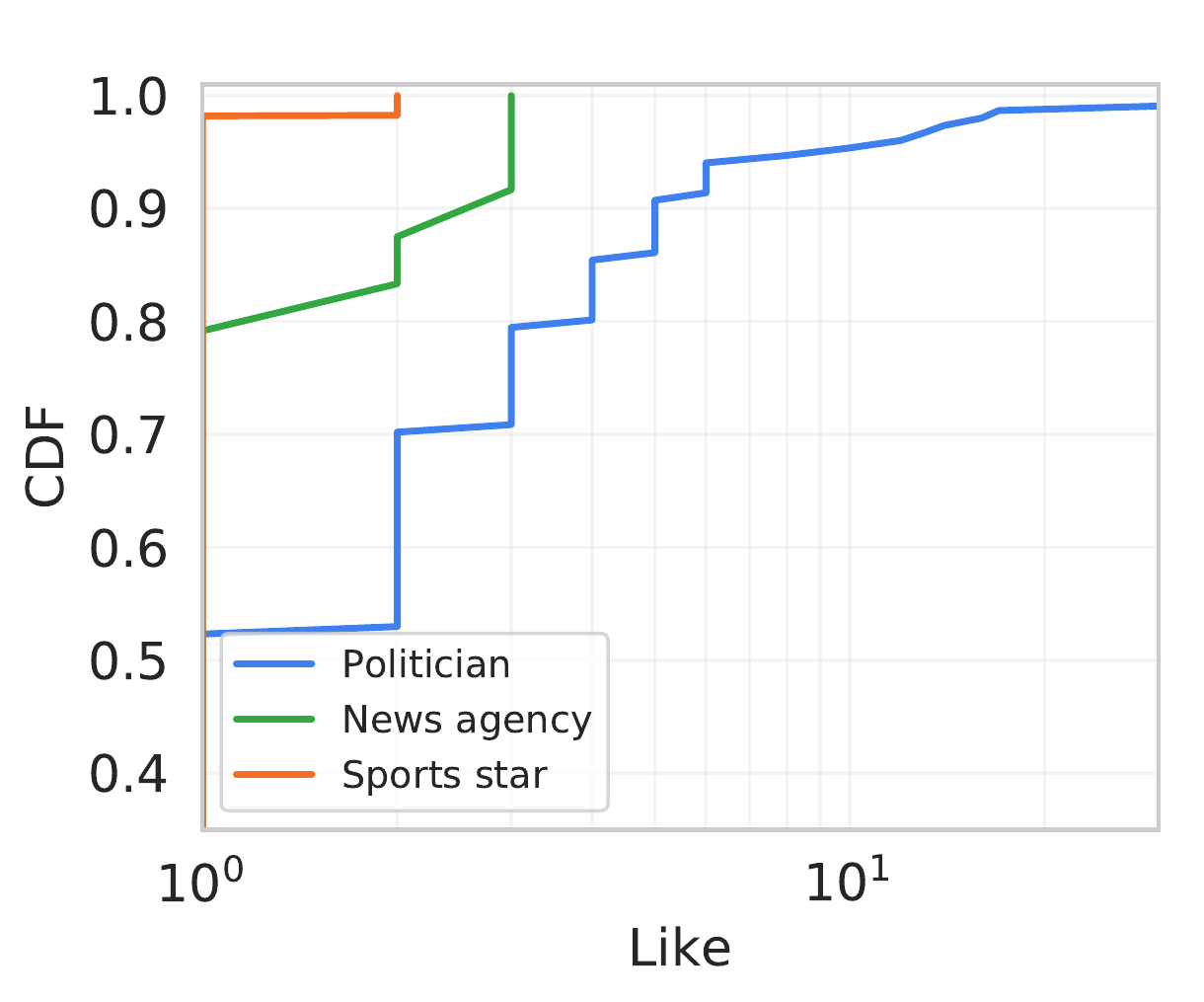}}\hfill
  \subfloat[Passive Engagement / cluster]{\includegraphics[width=0.49\linewidth, height=0.20\textwidth]{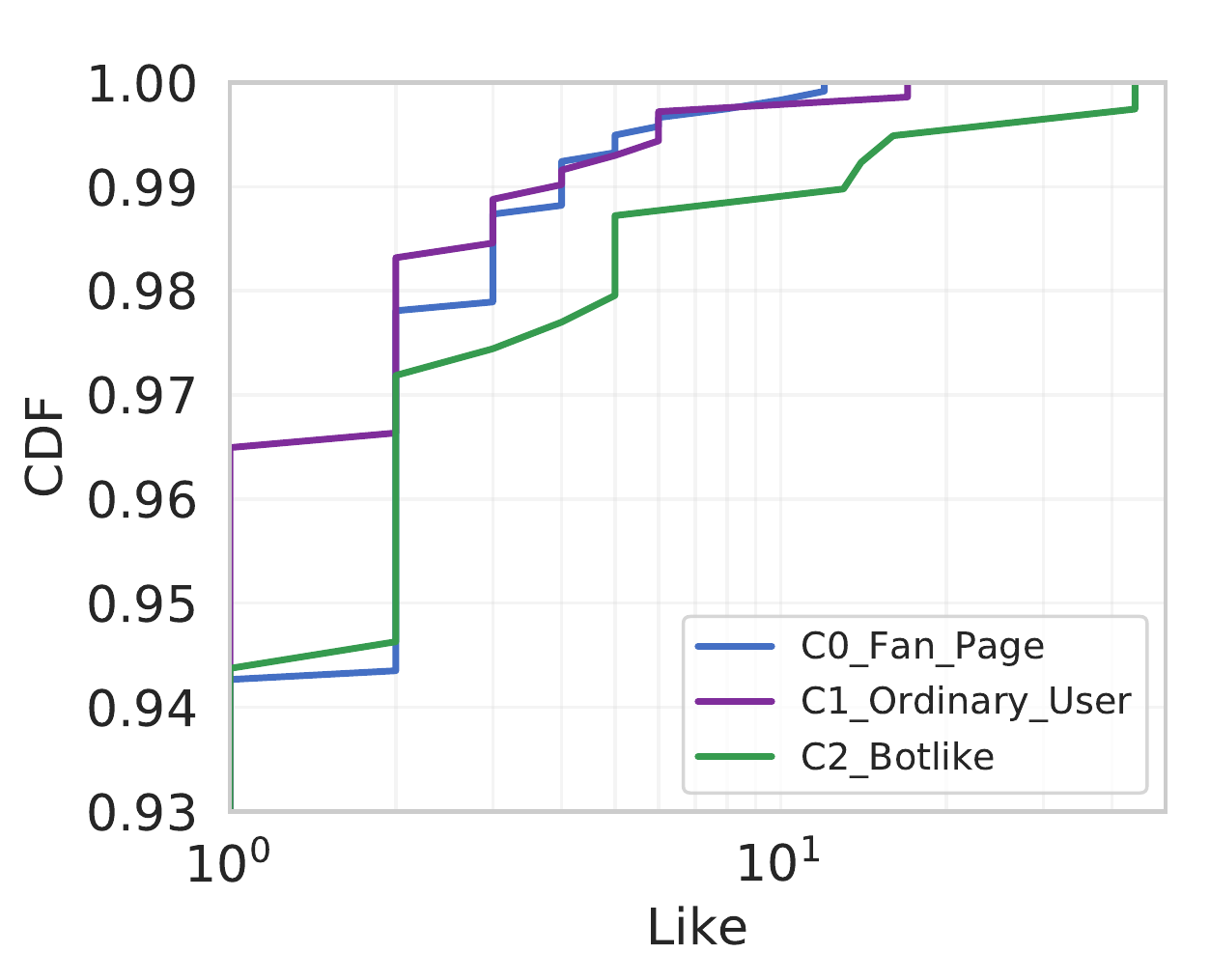}}%
 
\caption{Engagement per community and cluster: (a) CDF of number of comments issued by unique impersonator across communities (3 communities) and (b) across clusters (3 clusters). (c) CDF of number of likes issued by unique impersonator across communities and (d) across clusters.}
\label{fig_distribution_of_attention}
\vspace{-0.1cm}
\end{figure}


\section{What is the fake engagement of impersonators?}\label{behaviour_of_impersonator}

In this part, we move through the activity of impersonators to analyse how impersonators are distributing engagement and in general what is the rate of fake engagement through different communities?

\pb{Active \& Passive Engagement.}
First, let's look at the distribution of Active Engagement (comments) and Passive Engagement (likes) that are issued by impersonators across clusters and communities which is demonstrated in Figure \ref{fig_distribution_of_attention}. This figure displays the interest of impersonator amid communities/cluster. The first notable thing is that in Figure \ref{fig_distribution_of_attention}(a), while impersonators target all communities with a high number of comments, but politician and sports earn more (avg 4.89 vs. 2.81). This difference is even greater in passive engagement (Figure \ref{fig_distribution_of_attention}(c)) where sports star hosts the least number of likes compared to politicians (avg 1.01 vs. 2.75). Despite, the number of given comments in Sports star still high. Interestingly, across communities, impersonators mostly prefer to engage in the shape of Active Engagement rather than Passive Engagement.

Next, let's look at the distribution over clusters in Figure \ref{fig_distribution_of_attention}(b)(d). Again these engagements are given by unique impersonators. Interestingly, in all clusters, we can see impersonators issue more engagement in the shape of active engagement rather than passive. This shows the importance of content that is trying to publish. Moreover, as we expected, botlike is the most active cluster in promoting both active and passive engagements: While `C2\_Botlike' distributes more comments and likes, but `C0\_Fan\_Page' cluster issues fewer (avg. comments 5.9 vs. 2.29). 

\begin{figure}[ht]
\vspace{-0.3cm}
  \subfloat[Active Engagement]{\includegraphics[width=0.49\linewidth, , height=0.20\textwidth]{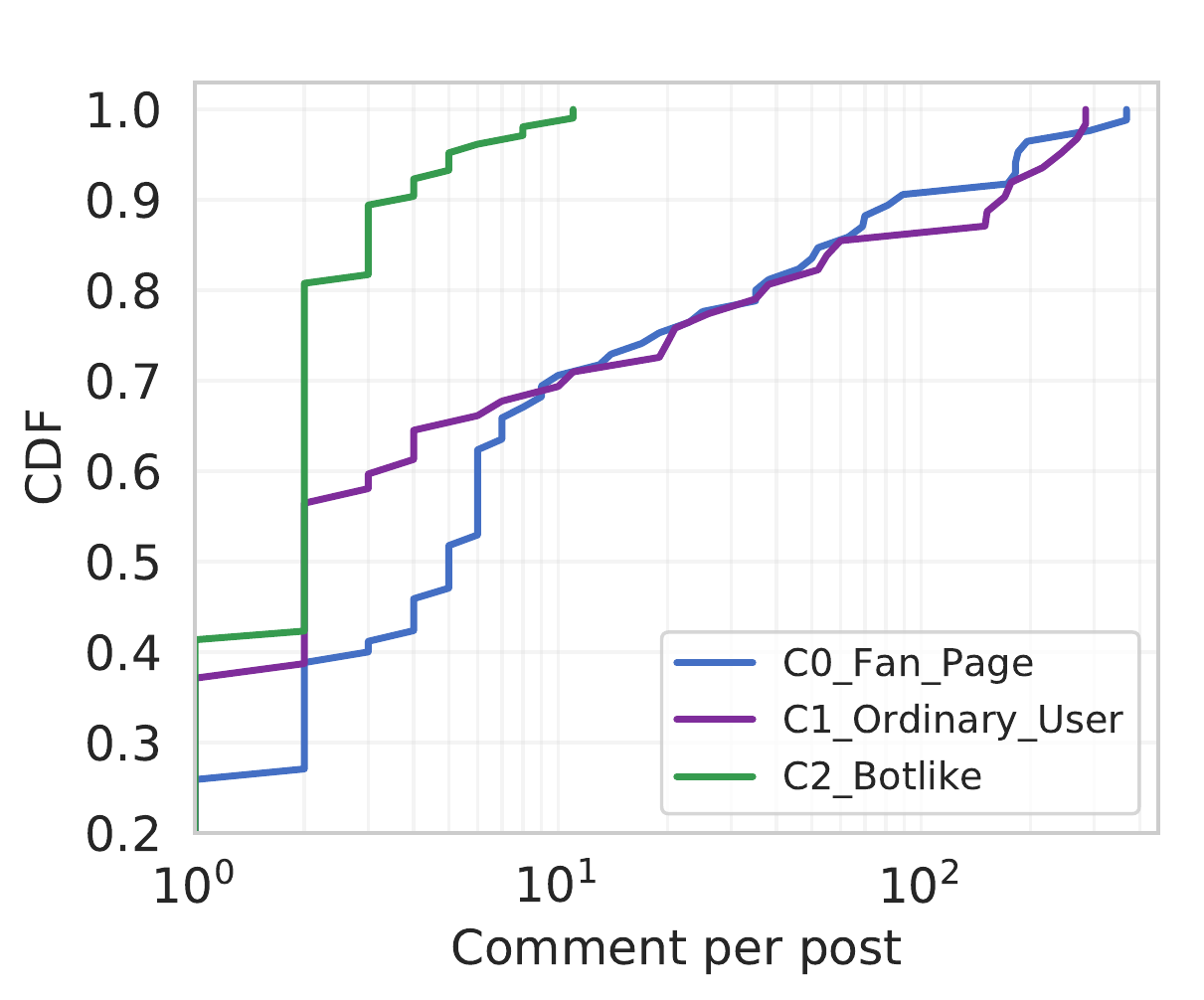}}\hfill
  \subfloat[Passive Engagement]{\includegraphics[width=0.49\linewidth, height=0.20\textwidth]{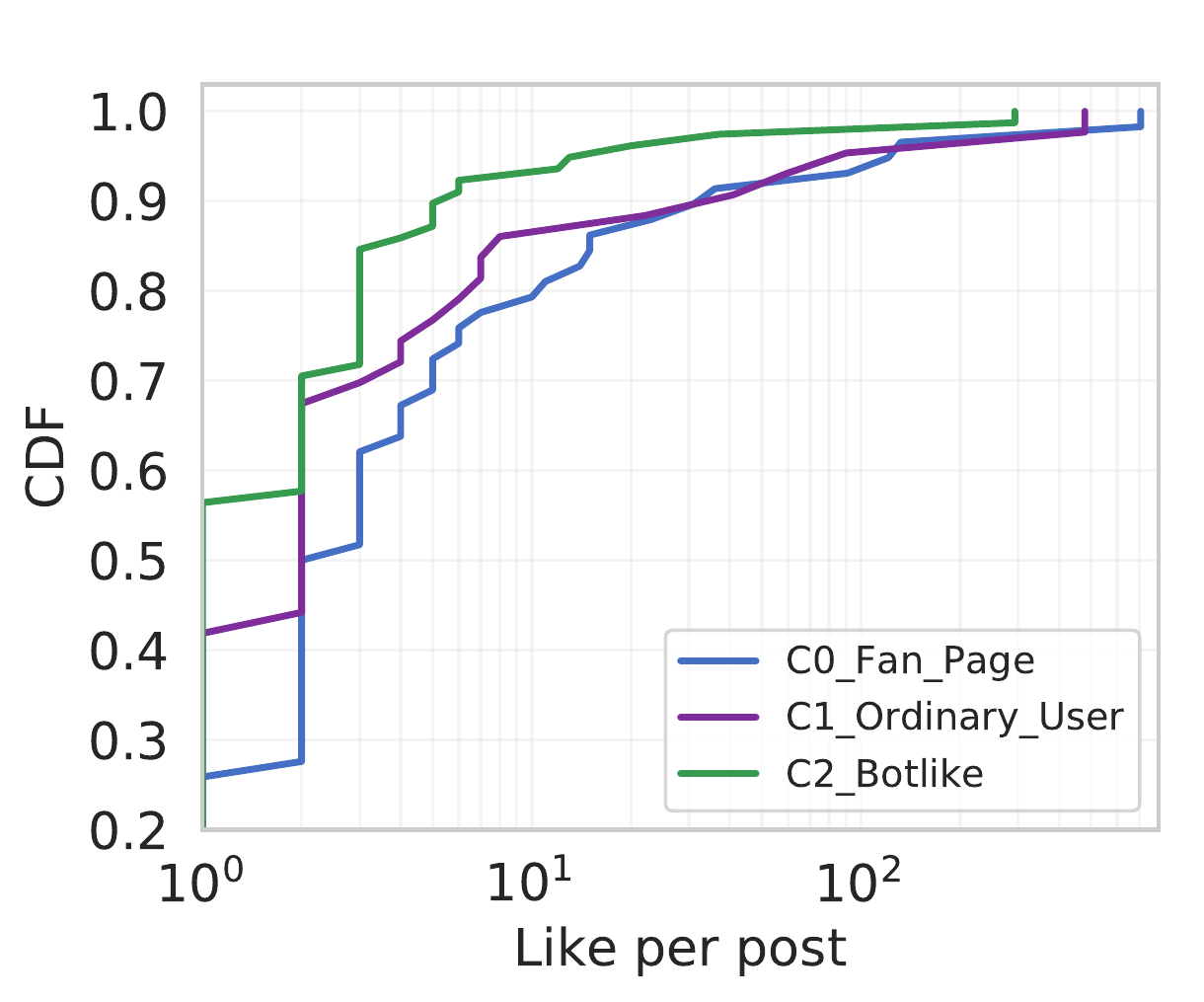}}%
\caption{Engagement per post: (a) CDF of number of comments issued by impersonator per post across clusters (b) CDF of number of likes issued by impersonator per post across clusters.}
\label{figure_comment_per_post}
\end{figure}

\pb{Post distribution.} Next, we examine posts to see how Active and Passive Engagements are scattered among them. This distribution is exhibited in Figure \ref{figure_comment_per_post}. On average, \textit{C0\_Fan\_Page} issued 29.5, \textit{C1\_Ordinary\_Users} issued 33.2, and \textit{C2\_Botlike} 2.04 comments per post. These numbers for like are 125.9, 260.8, and 19.3 per post respectively.

The first notable point is `C1\_Ordinary\_Users' spot more posts compared to `C2\_Botlike'. This behaviour is the same in both comments and like engagements. Interestingly, in comments, $\leq$80\% of `C2\_Botlike' cluster aim mostly 3 posts while `C0\_Fan\_Page' and `C1\_Ordinary\_User' target 10 times more posts (110). This reveals bots are targeting some specific (and limited) posts over communities to issue active engagement (comment), but are delivering like to all posts.

\begin{figure}[ht]
\vspace{-0.1cm}
  \subfloat[Communities]{\includegraphics[width=0.49\linewidth, , height=0.235\textwidth]{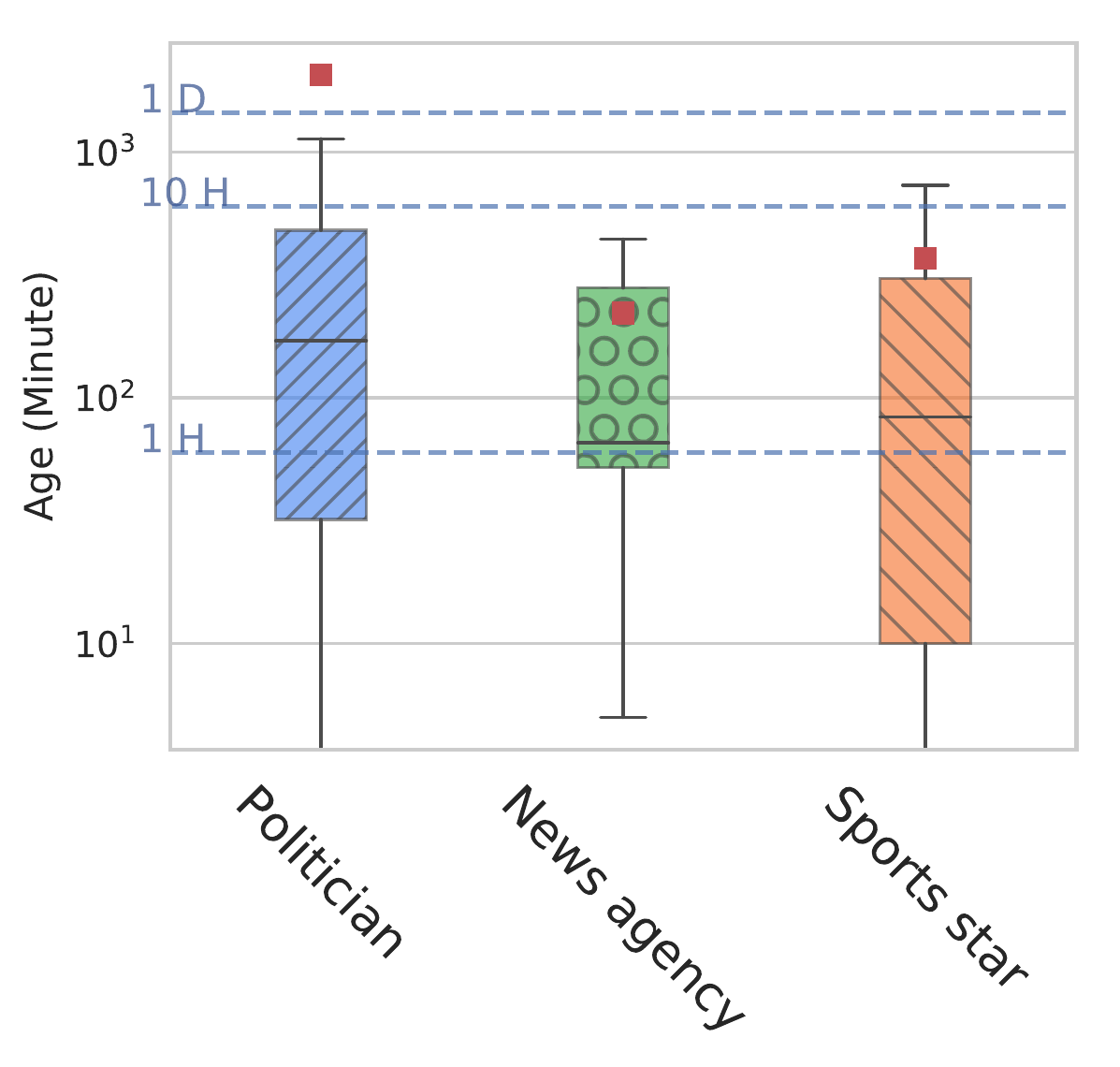}}\hfill
  \subfloat[Clusters]{\includegraphics[width=0.49\linewidth, height=0.24\textwidth]{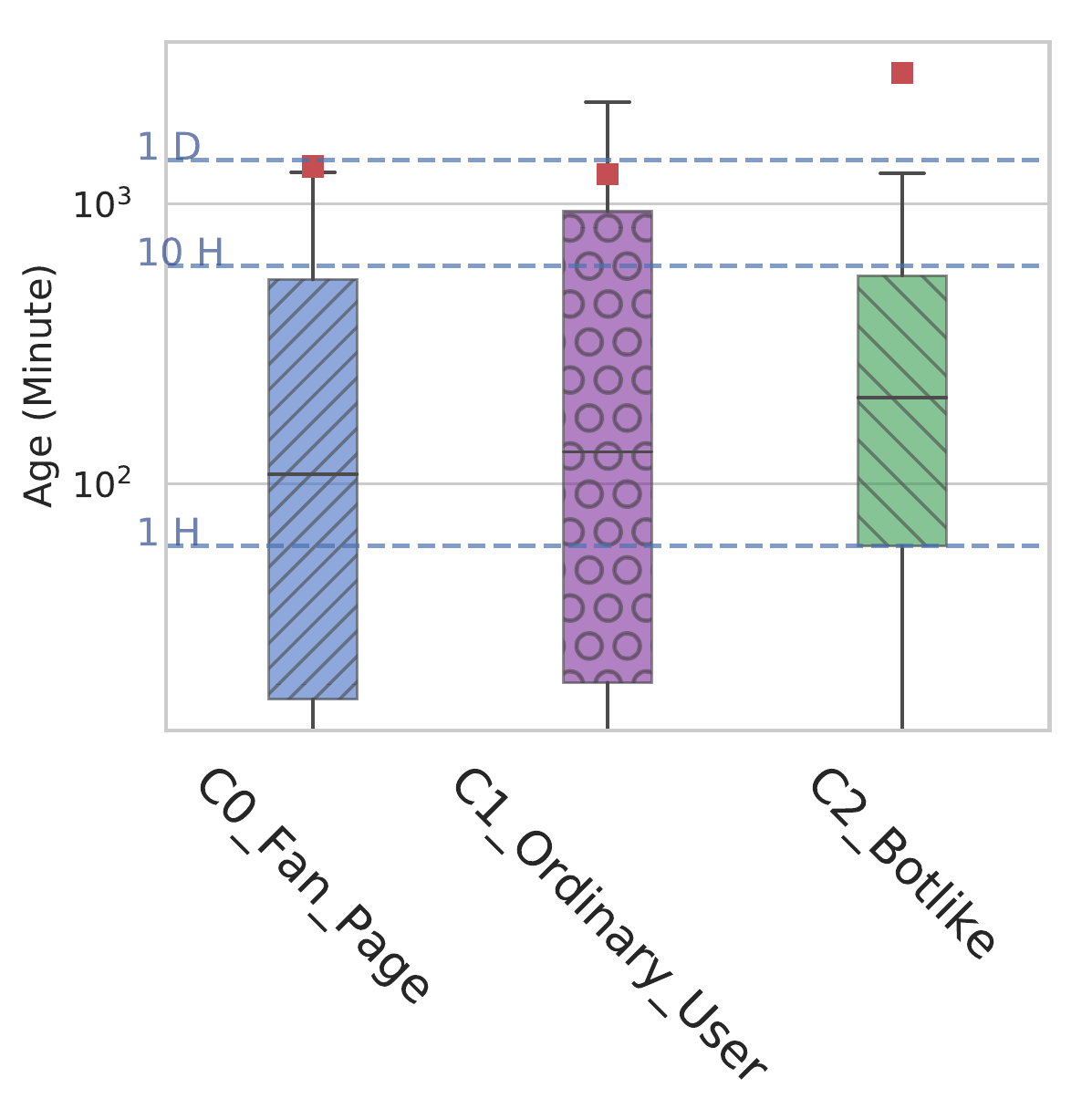}}%
\caption{ BoxPlot of Comment Age: (a) comment issued by impersonator across three communities.  (b) comment age issued by impersonator across three clusters.}
\label{plot_comment_age_community}
\vspace{-0.2cm}
\end{figure}

\pb{Comment Age.}
Another important metric which needs to be studied is the time of publishing comment by impersonators. This can support that the comments are published at the same time or not. This metric can be measured across both communities and clusters. From the community viewpoint, in Figure \ref{plot_comment_age_community}(a), impersonators publish sooner in sports star than politician communities (median 83.5 vs. 170). Moreover, the news agency has the shortest range of age (median 83.2) and impersonators mostly comment from hour $\geq$1. This number for the sports star is from minute $\geq$10.

From cluster viewpoint, Figure \ref{plot_comment_age_community}(b), both fan page (median 108) and ordinary user (median 130) clusters begin to write immediately. Moreover, both have the largest range. The interesting one is botlike cluster (median 203) which the range is between two fixed hours (1H to 10H). this behaviour reveals another peculiar characteristic of bots.

\section{What content do impersonator publish?}
\label{fake_content}

In this section, we aim to discuss what impersonators publish in form of comments? What is the content and topic of their comments? 
Is there any differences among clusters?

\begin{figure}[!ht]
\begin{center}
\vspace{-0.25cm}
  \subfloat{\includegraphics[width=1\linewidth]{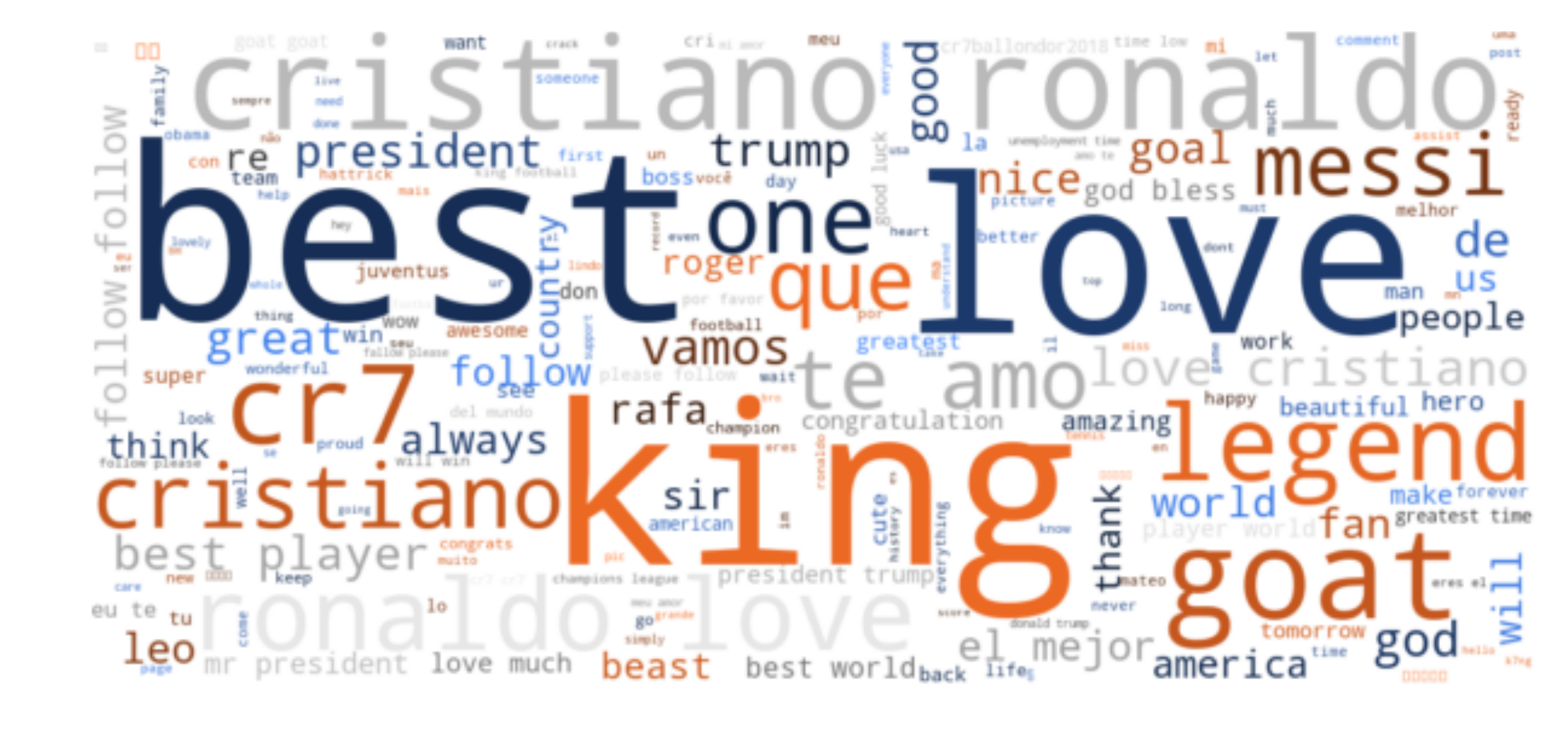}}\hfill \vspace{-0.1cm}
  \subfloat{\includegraphics[width=0.58\linewidth]{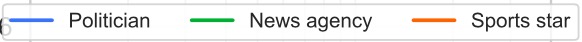}}%
 \vspace{-0.1cm}
 \end{center}
\caption{The most frequent words across communities.}
\label{plot_wordcloud_comment}
\vspace{-0.1cm}
\end{figure}

\pb{Analysis of Comments.}
First, let's look at the wordcloud of the most frequent word extracted from comments across communities (Figure \ref{plot_wordcloud_comment}). Words are colour-coded. Mostly, in Sports star community, we see `king' and `legend' keywords besides the name of the players which give positive support. Likewise, in politicians, we see the same trend and most dominant words are `best', `love', `great', and `thanks'. Moreover, some words such as `follow' (refers to follow me), and `story' (refers to check my story) are replicated which could be generated by bots that are mainly trying to attempt users to follow something.

\begin{figure}[ht]
\vspace{-0.2cm}
\centerline
{\includegraphics[width=0.46\textwidth, height=0.30\textwidth]{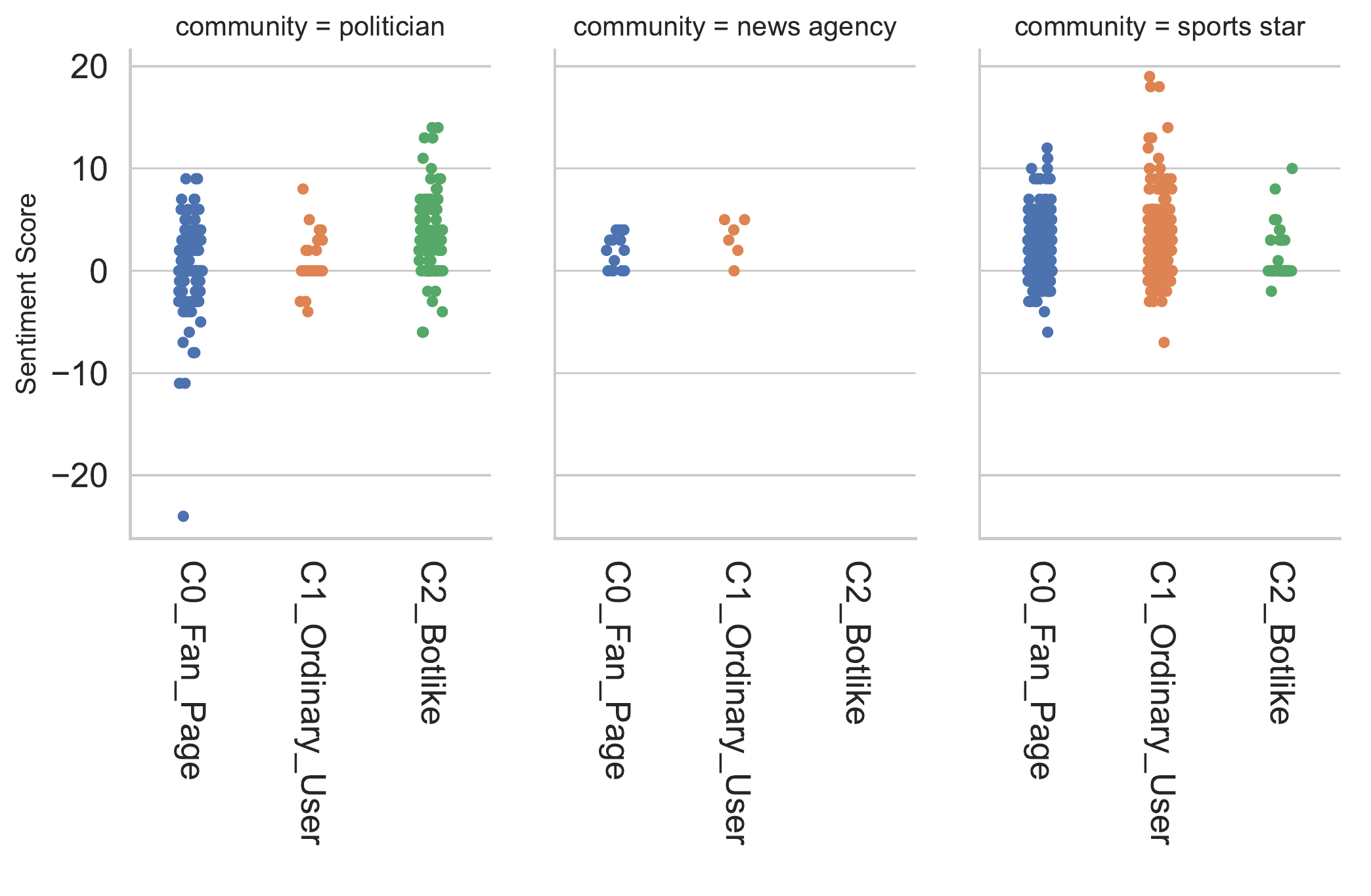}}
\vspace{-0.1cm}
\caption{Sentiment score for comments issued by impersonator across clusters and communities.}
\label{plot_sentiment_score}
\vspace{-0.1cm}
\end{figure}

Next, we aim to understand the general sentiment statistics across communities and clusters. This can give a good insight about the opinion expressed by impersonators. To do this, we use \textit{Affin} library \cite{nielsen2011new} which is one of the most popular lexicons that can be used. Figure \ref{plot_sentiment_score} presents general sentiment and summary statistics for comments distributed by impersonators in each cluster across communities. The output could be zero, positive, or negative number. In communities, we can see that the spread of sentiment polarity is much higher in sports star and politician as compared to news agency where a lot of the comments seem to be having a negative polarity. Practically, in the news agency, no negative comment exists. So, this community is a target for botlikes. However, we can observe a diverse range of comments in both politician and sports star communities. Form cluster viewpoint, the average sentiment for `C2\_Botlike' is 2, for `C1\_Ordinary\_User' is 0.7, and for `C0\_Fan\_Page' is 0.5. This claims bots distribute relatively positive text.

\begin{figure}[!ht]
  \subfloat[Per Cluster]{\includegraphics[width=0.39\linewidth, , height=0.21\textwidth]{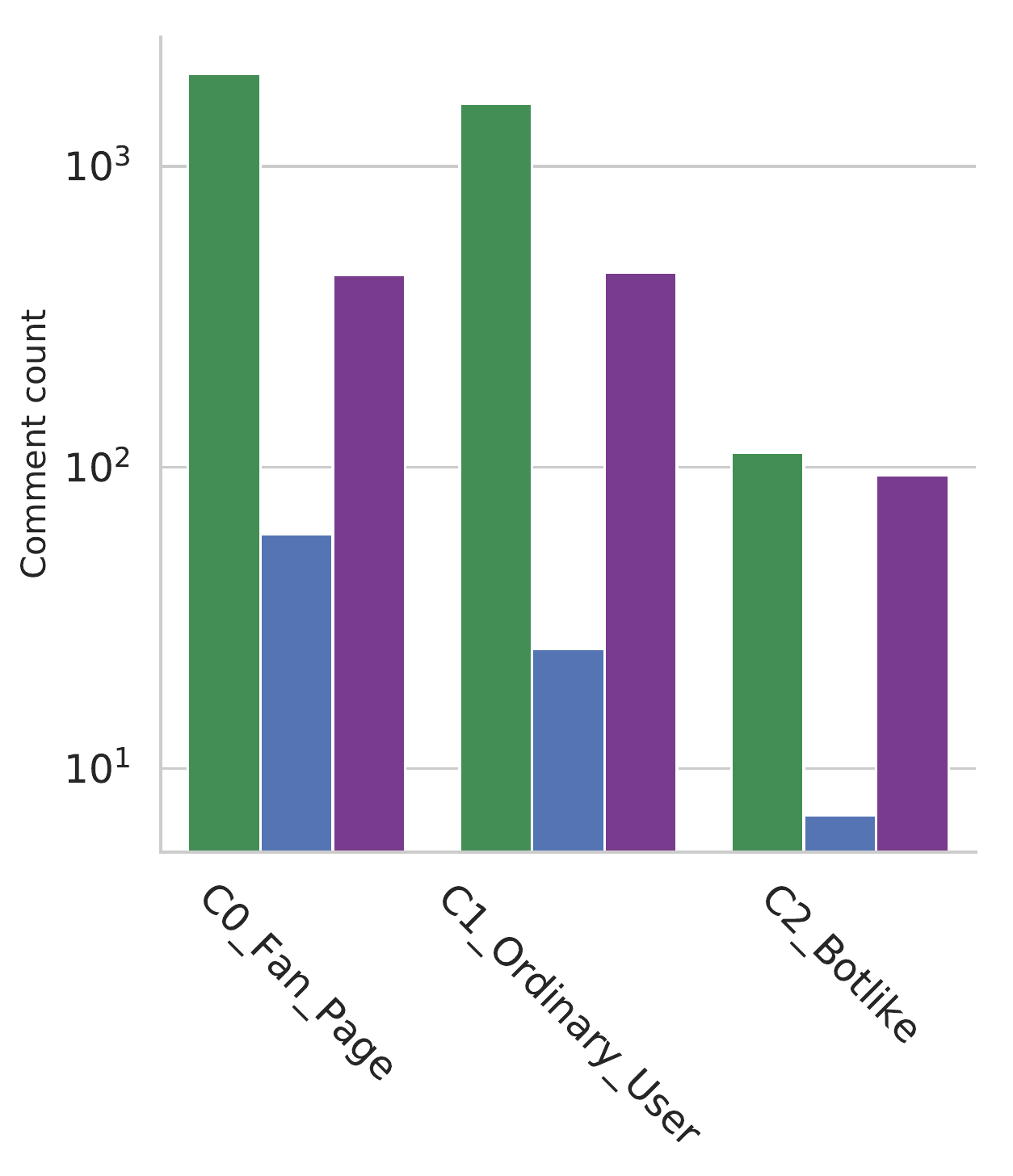}}\hfill
  \subfloat[Per Community]{\includegraphics[width=0.52\linewidth, height=0.21\textwidth]{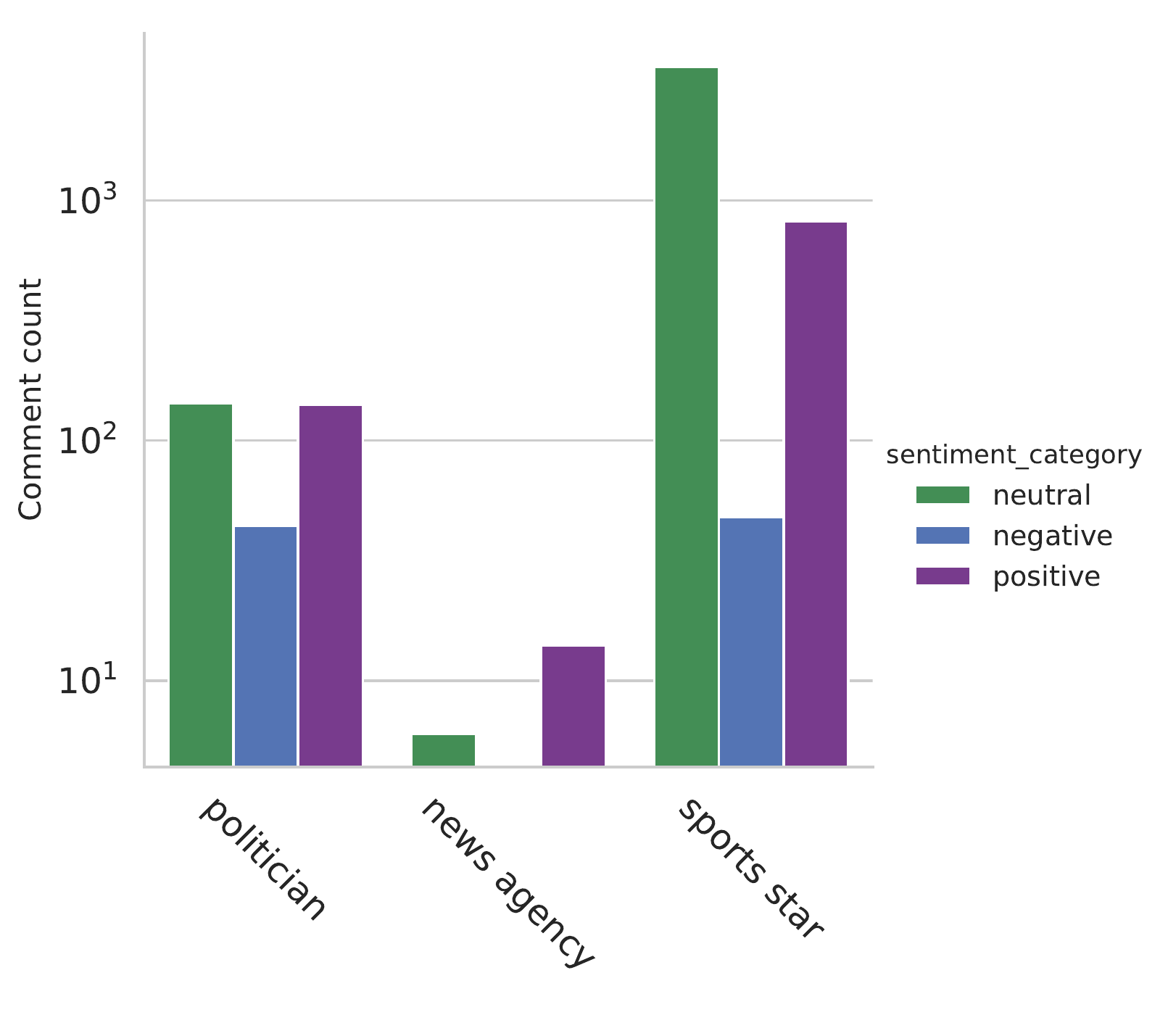}}%
\caption{Sentiment polarity: (a) Number of comment per cluster. (b) Number of comment per community.}
\label{fig_sentiment_count}
\end{figure}

Next, in Figure \ref{fig_sentiment_count} we visualize the frequency of sentiments across communities and clusters. Surprisingly, in Figure \ref{fig_sentiment_count}(a), while Fan pages published the most number of negative comments, but bots issued the least. Usually, fan pages are controlled by humans.  We list some random comments of impersonators across clusters in Table \ref{table_comment_examples}. \textit{``I post trump memes every day! Check out my page?"}; this comment (row [1]) caught in D. Trump post attempting audience to follow fan pages. These kind of comments are repeated over different posts and clearly are published by bots.

In Figure \ref{fig_sentiment_count}(b), by considering communities, Sports star have more neutral comments due to the presence of comments which are talking about sporting events without the presence of any emotion or feeling. Besides, both sports star and politician have the same rate of negative comments. This is an example of positive comment in politicians (Table \ref{table_comment_examples}, row [3]): \textit{``You are my president and I love you forever"}, and this is an example of negative in sports star (Table \ref{table_comment_examples}, row [16]): \textit{``kill ur self"}.

\begin{figure}[ht]
\vspace{-0.25cm}
  \subfloat[Duplication per Cluster]{\includegraphics[width=0.49\linewidth, , height=0.20\textwidth]{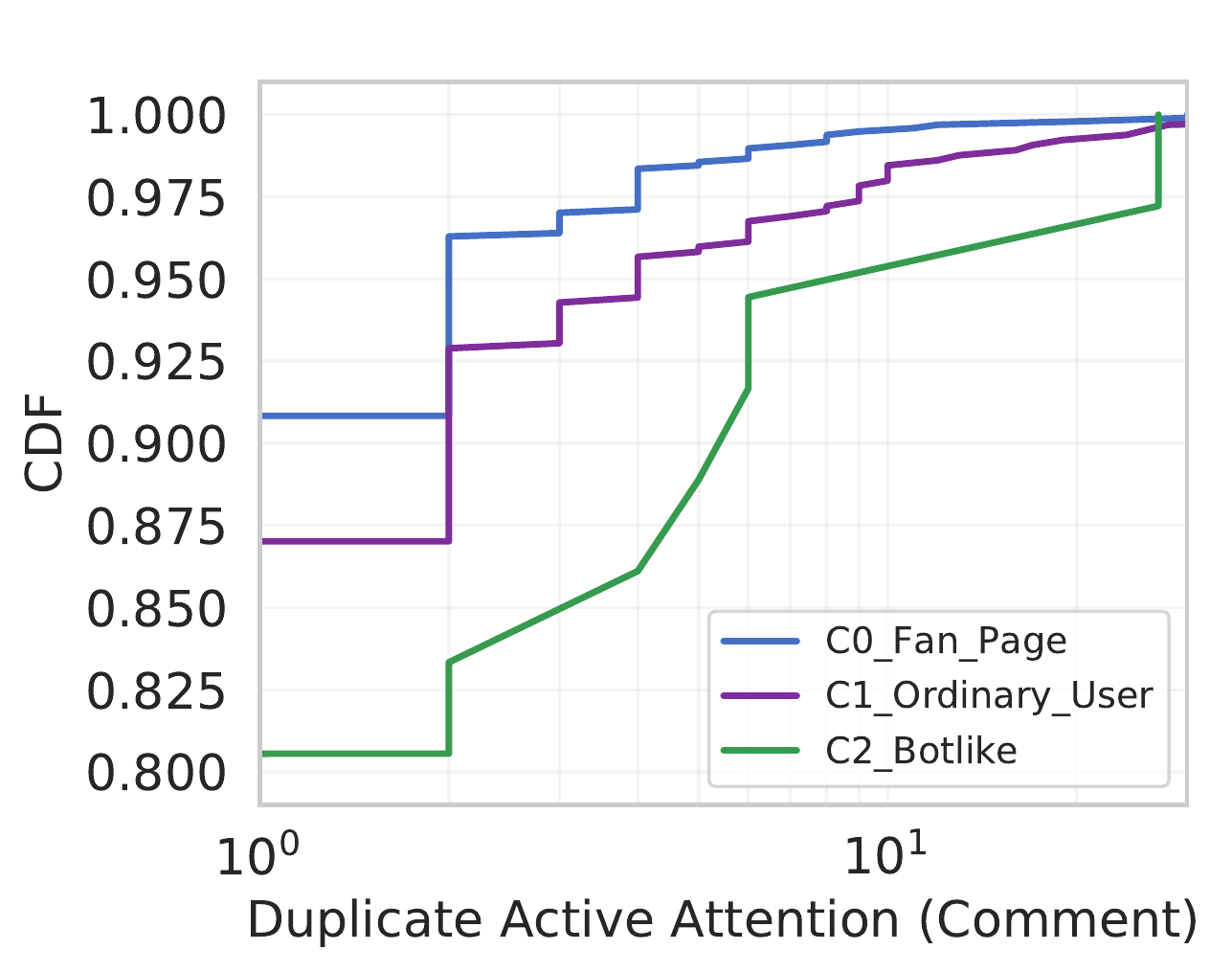}}\hfill
  \subfloat[Duplication per Community]{\includegraphics[width=0.49\linewidth, height=0.20\textwidth]{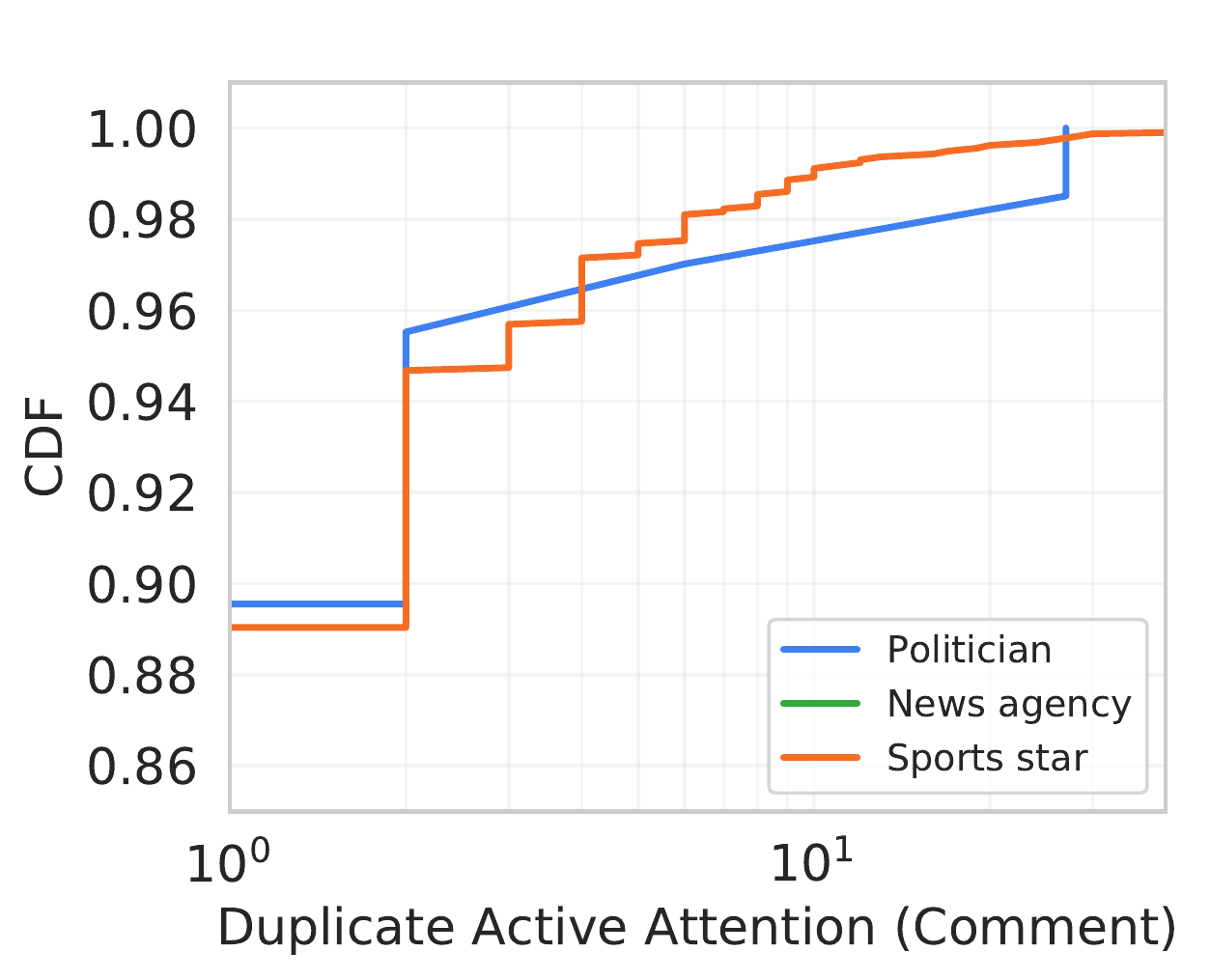}}%
\caption{CDF of number of duplicated comments: (a)   across clusters  (b) across communities.}
\label{figure_duplicated_commet}
\vspace{-0.1cm}
\end{figure}

\pb{Duplication of comments.}
Next, we investigate how many duplicated comments are published by impersonators. This important metric confirms if clusters follow some particular patterns of publishing or hire automated bots to advertise something with a frequency. Toward that end, we implemented a similarity module and we were able to identify duplicated comments across communities which are demonstrated in Figure \ref{figure_duplicated_commet}. 
To estimate the degree of similarity between comments, we use cosine similarity technique and we employ \textit{NLTK} \cite{Loper02nltk:the} library inside \textit{scikit-learn} \cite{scikit-learn}. Comments that have a similarity of $\geq$0.95 (scale 0 to 1) are considered as duplicated text. Note that emojis are skipped from the measurement. As we observed, all clusters are utilising pre-defined text (with high rate of similarity) and common patterns to publish comments. It is crystal clear in Figure \ref{figure_duplicated_commet}(a), \textit{botlike} cluster distributes with a higher unusual rate (median 7) than ordinary users (median 3) and fan pages (median 2). To be sure that duplicated comments are distributed with pre-defined algorithms, we manually checked the comment time. $\geq$95\% of the duplicated comments have the same publishing age (from the post) which clarify our claim. 

From community viewpoint, in Figure \ref{figure_duplicated_commet}(b), impersonators hugely target politician (median 2) and sports star (median 2) communities by writing repeated comments and surprisingly there is no sign of duplication in the news agency. We, next check the text of duplicated comments by manual examination among all. 

This part contains more experiments on the text and the correlation of the words that are used by clusters. As the matter of space, we include other plots in a separated report available online\footnote{https://sites.google.com/view/iengagement/home}.
In general, $\geq$75\% of C0\_Fan\_Page duplicated comments are emojis, and hugely they invite audience by pre-defined text such as ``follow us" and ``best of Ronaldo" to their pages to gain followers. In C1\_Ordinary\_Users, again we see $\leq$53\% of comments are emojis. Comments contain human-generated text and related hashtags are used to express the support. Most of the emojis are positive such as ``heart", ``like", and ``thumbs up". C2\_Botlike cluster contains $\leq$\%70 emojis in both positive and negative feeling. Comments hugely hold very short text, hashtags, and sometimes mentions (start with @ sign).

\begin{table}[ht]
\vspace{0.2cm}
\caption{Some example of comments}
\begin{center}
\vspace{-0.3cm}
\resizebox{\columnwidth}{!}{
 \scalebox{1.2}{
\begin{tabular}{|c|c|c|}
\hline
\textbf{\#} & \textbf{Cluster} & \textbf{written comments by impersonators (randomly selected)} \\ \hline
\textbf{1} & \multirow{6}{*}{\textbf{C0\_Fan\_Page}} & \textit{I post trump memes every day! Check out my page?} \\ \cline{1-1} \cline{3-3} 
\textbf{2} &  & \textit{king leo} \\ \cline{1-1} \cline{3-3} 
\textbf{3} &  & \textit{You are my president and I love you forever} \\ \cline{1-1} \cline{3-3} 
\textbf{4} &  & \textit{Thanks for supporting ALL Americans!} \\ \cline{1-1} \cline{3-3} 
\textbf{5} &  & \textit{if u love messi like this comment} \\ \cline{1-1} \cline{3-3} 
\textbf{6} &  & \textit{Follow @mtfoot for more!} \\ \hline
\textbf{7} & \multirow{6}{*}{\textbf{C1\_Ordinary\_User}} & \textit{Congratulations juventus team and @cristiano} \\ \cline{1-1} \cline{3-3} 
\textbf{8} &  & \textit{King cristiano} \\ \cline{1-1} \cline{3-3} 
\textbf{9} &  & \textit{\begin{tabular}[c]{@{}c@{}}We are with you, if you not gone win the balondor, \\ will be back stronger next year! forzaaa @cristiano Te Amo\end{tabular}} \\ \cline{1-1} \cline{3-3} 
\textbf{10} &  & \textit{You deserve all the best next time you want the hat-trick} \\ \cline{1-1} \cline{3-3} 
\textbf{11} &  & \textit{Good luck legend, I hope you scored 3 goals tomorrow} \\ \cline{1-1} \cline{3-3} 
\textbf{12} &  & Great win for @juventus, so happy @cristiano \\ \hline
\textbf{13} & \multirow{6}{*}{\textbf{C2\_Botlike}} & \textit{\begin{tabular}[c]{@{}c@{}}More Americans are now employed than ever \\ recorded before in our nation's history. President Donald Trump\end{tabular}} \\ \cline{1-1} \cline{3-3} 
\textbf{14} &  & \textit{President Trump miracle from God \& for the country} \\ \cline{1-1} \cline{3-3} 
\textbf{15} &  & \textit{Check my profile and my story} \\ \cline{1-1} \cline{3-3} 
\textbf{16} &  & \textit{kill ur self} \\ \cline{1-1} \cline{3-3} 
\textbf{17} &  & \textit{Beautiful pics Rafa! Thank you for sharing} \\ \cline{1-1} \cline{3-3} 
\textbf{18} &  & thanks God that you are our president \\ \hline

 \multicolumn{2}{ c }
{Note: Emojis are removed from the text.}

\end{tabular}
}}
\end{center}
\label{table_comment_examples}
\vspace{-0.3cm}
\end{table}



\section{Conclusion and Future Work}\label{future}

\pb{In conclusion,} this paper has performed a first analysis of the content and engagements generated by impersonators on Instagram. Based on our previous studies, we did an investigation to discover the behaviour of impersonators and the generated content in three major communities. To the best of our knowledge, this is the first paper that conducts such analysis on Instagram. We used the dataset of nearly 4K impersonator which is extracted from our previous paper \cite{Zare1910:Typification}. We analysed the distribution of issued Active and Passive engagement given by impersonator across three major communities. Next, we focused to the written comments by impersonators to perceive what kind of content do they publish. We obtained valuable knowledge by using various text analysis techniques which explains better the behaviours of impersonators. 

\pb{As future work,} This study could be extended from various angles: first, it is desirable to train a machine/deep learning classifier for comments. This can be done by considering some other important profile metrics alongside text features. As a result, this model could predict at first whether the content of the text is fake or not and second, evaluate whether the publisher of that comment is impersonator or not. Another perspective is to study other social media to understand is there any similar pattern across different platforms and can we correlate the identified profiles of impersonators.


\bibliographystyle{unsrt}
\bibliography{conference_041818}

\begin{thebibliography}{10}

\bibitem{Zare1910:Typification}
Koosha Zarei, Reza Farahbakhsh, and Noel Crespi.
\newblock Typification of impersonated accounts on instagram.
\newblock In {\em 2019 IEEE 38th International Performance Computing and
  Communications Conference (IPCCC) (IPCCC 2019)}, London, United Kingdom
  (Great Britain), October 2019.

\bibitem{kooshaDeep}
Koosha Zarei, Reza Farahbakhsh, and Noel Crespi.
\newblock Deep dive on politician impersonating accounts in social media.
\newblock In {\em 2019 IEEE Symposium on Computers and Communications (ISCC)
  (IEEE ISCC 2019)}, Barcelona, Spain, June 2019.

\bibitem{impersonaion_law}
https://definitions.uslegal.com/c/criminal-impersonation/, 2019.

\bibitem{8622011}
L.~{Caruccio}, D.~{Desiato}, and G.~{Polese}.
\newblock Fake account identification in social networks.
\newblock In {\em 2018 IEEE International Conference on Big Data (Big Data)},
  pages 5078--5085, Dec 2018.

\bibitem{RAMALINGAM2018165}
Devakunchari Ramalingam and Valliyammai Chinnaiah.
\newblock Fake profile detection techniques in large-scale online social
  networks: A comprehensive review.
\newblock {\em Computers \& Electrical Engineering}, 65:165 -- 177, 2018.

\bibitem{Ferrara:2016:RSB:2963119.2818717}
Emilio Ferrara, Onur Varol, Clayton Davis, Filippo Menczer, and Alessandro
  Flammini.
\newblock The rise of social bots.
\newblock {\em Commun. ACM}, 59(7), 2016.

\bibitem{profchar}
White J. S. Hudson B. Voter B. R. \& Matthews J.~N. Gurajala, S.
\newblock Profile characteristics of fake twitter accounts.
\newblock 2016.

\bibitem{7865975}
S.~Shehnepoor, M.~Salehi, R.~Farahbakhsh, and N.~Crespi.
\newblock Netspam: A network-based spam detection framework for reviews in
  online social media.
\newblock {\em IEEE Transactions on Information Forensics and Security},
  12(7):1585--1595, July 2017.

\bibitem{Xiao:2015:DCF:2808769.2808779}
Cao Xiao, David~Mandell Freeman, and Theodore Hwa.
\newblock Detecting clusters of fake accounts in online social networks.
\newblock In {\em Proceedings of the 8th ACM Workshop on Artificial
  Intelligence and Security}, AISec '15, pages 91--101. ACM, 2015.

\bibitem{stweeler}
Zafar Gilani, Reza Farahbakhsh, Gareth Tyson, Liang Wang, and Jon Crowcroft.
\newblock Of bots and humans (on twitter).
\newblock In {\em Proceedings of the 2017 IEEE/ACM International Conference on
  Advances in Social Networks Analysis and Mining 2017}, ASONAM '17, pages
  349--354, New York, NY, USA, 2017. ACM.

\bibitem{tweb19}
Zafar Gilani, Reza Farahbakhsh, Gareth Tyson, and Jon Crowcroft.
\newblock A large-scale behavioural analysis of bots and humans on twitter.
\newblock {\em ACM Trans. Web}, 13(1):7:1--7:23, February 2019.

\bibitem{Li_2016_WCC}
Yixuan Li, Oscar Martinez, Xing Chen, Yi~Li, and John~E. Hopcroft.
\newblock In a world that counts: Clustering and detecting fake social
  engagement at scale.
\newblock In {\em Proceedings of the 25th International Conference on World
  Wide Web}, WWW '16, pages 111--120, Republic and Canton of Geneva,
  Switzerland, 2016. International World Wide Web Conferences Steering
  Committee.

\bibitem{Sen:2018:WWL:3201064.3201105}
Indira Sen, Anupama Aggarwal, Shiven Mian, Siddharth Singh, Ponnurangam
  Kumaraguru, and Anwitaman Datta.
\newblock Worth its weight in likes: Towards detecting fake likes on instagram.
\newblock In {\em Proceedings of the 10th ACM Conference on Web Science},
  WebSci '18. ACM, 2018.

\bibitem{Buccafurri:2015:CTF:2822539.2822620}
Francesco Buccafurri, Gianluca Lax, Serena Nicolazzo, and Antonino Nocera.
\newblock Comparing twitter and facebook user behavior.
\newblock {\em Comput. Hum. Behav.}, 52(C):87--95, November 2015.

\bibitem{Lim:2015:MVI:2808797.2808820}
Bang~Hui Lim, Dongyuan Lu, Tao Chen, and Min-Yen Kan.
\newblock Mytweet via instagram: Exploring user behaviour across multiple
  social networks.
\newblock IEEE/ACM ASONAM '15, pages 113--120. ACM, 2015.

\bibitem{Choumane:2017:PMS:3093241.3093258}
Ali Choumane, Zein Al~Abidin~Ibrahim, and Bilal Chebaro.
\newblock Profiles matching in social networks based on semantic similarities
  and common relationships.
\newblock In {\em Proceedings of the International Conference on Compute and
  Data Analysis}, ICCDA '17, pages 14--18. ACM, 2017.

\bibitem{Krombholz2012}
Katharina Krombholz, Dieter Merkl, and Edgar Weippl.
\newblock Fake identities in social media: A case study on the sustainability
  of the facebook business model.
\newblock {\em Journal of Service Science Research}, 4(2), Dec 2012.

\bibitem{Goga:2015:RPM:2783258.2788601}
Oana Goga, Patrick Loiseau, Robin Sommer, Renata Teixeira, and Krishna~P.
  Gummadi.
\newblock On the reliability of profile matching across large online social
  networks.
\newblock In {\em Proceedings of the 21th ACM SIGKDD International Conference
  on Knowledge Discovery and Data Mining}, KDD '15, pages 1799--1808. ACM,
  2015.

\bibitem{InstagramAPI:online}
Instagram.
\newblock Official api graph instagram, September 2019.

\bibitem{InstagramBadge}
Instagram~Verified Badges.
\newblock Instagram verified badges, September 2019.

\bibitem{nielsen2011new}
Finn Årup Nielsen.
\newblock A new anew: Evaluation of a word list for sentiment analysis in
  microblogs, 2011.

\bibitem{Loper02nltk:the}
Edward Loper and Steven Bird.
\newblock Nltk: The natural language toolkit.
\newblock In {\em In Proceedings of the ACL Workshop on Effective Tools and
  Methodologies for Teaching Natural Language Processing and Computational
  Linguistics. Philadelphia: Association for Computational Linguistics}, 2002.

\bibitem{scikit-learn}
F.~Pedregosa, G.~Varoquaux, and A.~et~al. Gramfort.
\newblock {Scikit-learn: Machine Learning in Python }.
\newblock {\em Journal of Machine Learning Research}, 12:2825--2830, 2011.

\end{thebibliography}

\end{document}